# Elastomeric Nematic Colloids, Colloidal Crystals and Microstructures with Complex Topology

Ye Yuan,[a] Patrick Keller,[b] and Ivan I. Smalyukh*[a,c,d]

[a]Department of Physics, University of Colorado, Boulder, CO 80309, USA

[b]Laboratoire Physico-Chimie Curie, Institut Curie, PSL Research University, CNRS UMR 168, F-75248 Paris, France

[c]Department of Electrical, Computer, and Energy Engineering, Materials Science and Engineering Program and Soft Materials Research Center, University of Colorado, Boulder, CO 80309, USA

[d]Renewable and Sustainable Energy Institute, National Renewable Energy Laboratory and University of Colorado, Boulder, CO 80309, USA

*Correspondence to: ivan.smalyukh@colorado.edu

† Electronic supplementary information (ESI) available.

Control of physical behaviors of nematic colloids and colloidal crystals has been demonstrated by tuning particle shape, topology, chirality and surface charging. However, the capability of altering physical behaviors of such soft matter systems by changing particle shape and the ensuing responses to external stimuli has remained elusive. We fabricated genus-one nematic elastomeric colloidal ring-shaped particles and various microstructures using two-photon photopolymerization. Nematic ordering within both the nano-printed particle and the surrounding medium leads to anisotropic responses and actuation when heated. With the thermal control, elastomeric microstructures are capable of changing from genus-one to genus-zero surface topology. Using these particles as building blocks, we investigated elastomeric colloidal crystals immersed within a liquid crystal fluid, which exhibit crystallographic symmetry transformations. Our findings may lead to colloidal crystals responsive to a large variety of external stimuli, including electric fields and light. Pre-designed response of elastomeric nematic colloids, including changes of colloidal surface topology and lattice symmetry, are of interest for both fundamental research and applications.

1. Introduction

Colloids are abundant, attracting interest of physicists, biologists, chemists and engineers alike [1-5]. Examples of colloids range from milk to quantum dots and from gold nanoparticles to many food and personal care products [1-3]. At the same time, colloidal interactions promise inexpensive scalable means of fabricating mesostructured materials from nanoparticles by self-assembly. Colloidal interactions in liquid crystals (LCs) are particularly interesting and exhibit contributions due to the minimization of energetic costs of director deformations induced by particles, which stem from the orientational elasticity of the LC host medium [4-20]. A sub-field of soft matter at the interfaces of LC and colloidal sciences, these systems bridge the LC and colloidal sciences with fields as diverse as topology [21-28], biology and photonics [29-35]. One of the key research goals is to develop a means for controlling such nematic colloidal behavior by external stimuli, which could potentially also lead to technological applications [36-49]. However, while shapes and topology of colloidal inclusions were shown to be capable of defining formation of topological defects and inter-particle colloidal interactions [8,16,23], so far only theoretical studies considered the possibility of transformations of shape and topological characteristics like genus within such colloidal systems [50]. Similar to nematic LC colloids, controllable actuation of components within composite mesostructured materials are needed in photonic crystals and other microstructures, but progress is limited so far. The recent advances in developing LC elastomeric structures [51-63] still remain to be extended to colloidal systems and microstructures of similar dimensions, even though elastomers were used to fabricate reconfigurable colloidal objects with soft-lithography [61,62].

In this study, we describe genus-one nematic elastomeric colloidal rings and similar-shaped microstructures fabricated using two-photon photopolymerization and immersed in an unpolymerized surrounding nematic host of the same nematic precursors. Thermal stimuli prompt anisotropic actuations of shape and changes of genus surface characteristics, as well as changes of symmetry and re-configuration of two-dimensional (2D) lattices they form. Our work demonstrates how elastomeric nematic colloids can be used in probing the fundamental role of topology and geometry in nematic colloids, and how the interaction of elastomeric microstructures and fluid nematic host medium could potentially enrich electro-optic and photonic applications of LCs and elastomers.

## 2. Experimental

The LC elastomeric (LCE) monomers used in this study are synthesized following established protocols reported previously [63]. Other constituents of the LCE mixture, e.g., the photoinitiator and crosslinker are purchased from Sigma Aldrich and used without further purification (Fig. 1). The chemicals are first dissolved in dichloromethane and then vacuum dried to produce a homogeneous mixture, which are then infiltrated in liquid phase between a glass slide and a cover slip at 85 °C. The inner surfaces of confining glass substrates are spin-coated with PI2555 (HD MicroSystem) or SE5661 (Nissan Chemical) to produce planar (rubbed unidirectionally) or homeotropic (perpendicular) surface anchoring boundary conditions to define the far-field LC molecular orientation $\mathbf{n}_0$. The gap between the glass slide and the coverslip is defined by thin films of 20-50 μm thick. Shielded from direct exposure to ambient light, the sample is cooled down to room temperature (25 °C) at a rate of 0.5 °C/min; we then let it stay overnight to produce a well-aligned LCE mixture in nematic phase over large area. By mixing the two types of monomers, we obtain the mixture in nematic phase at a temperature much lower than that when they are used alone [63]. This enlarged thermal range of nematic stability makes such mixtures more suitable for the use in two-photon photopolymerization based fabrication of colloidal particles.

The elastomeric colloidal rings and microstructures are fabricated within the obtained nematic mixture using a homemade 3D microfabrication system [25] based on two-photon photopolymerization (Fig. 2a). A femtosecond laser (Chameleon Ultra, from Coherent) operating at 780 nm polymerizes the mixture at the focal point of an objective lens (Olympus UPLFLN 40×, NA=0.75); the laser beam's power is measured to be 15-20 mW before entering the objective. A 3D nanopositioning stage (Physik Instrumente, NanoCube P-611.3SF) controls the movement of the sample with trajectories defined by computer programs; the "printing" speed is about 20 μm s$^{-1}$. The same program also controls the laser by opening and closing a fast shutter (Uniblitz, LS3Z2). Unless noted otherwise, the rings are printed within the bulk of the rather viscous mixture, not connected to the glass substrates, to allow free movement and actuation without constraints (Fig. S1, ESI†). After the polymerization, the sample is tested as-is, or washed with isopropanol to remove the unpolymerized monomer mixture. To strengthen and support the LCE structures, a scaffold made of an isotropic polymer (NanoScribe, IP-L) can be "printed" first and then the LCE structures incorporating them can be additively incorporated. The thermal response tests are performed on a temperature controlling stage (Instec HCS302XY) mounted on a microscope (Olympus BX51). When unpolymerized mixtures are present, both printing and testing are done under red light to avoid unintended polymerization of the sample.

## 3. Results & discussion

### 3.1 Shape change and thermal actuation of elastomeric colloidal rings

Maintaining the shape of printed colloidal structures in the monomer mixture can be a challenge. Particles shaped as lines or circles produced in this condition may curl and distort significantly after printing without external stimuli [55] (Fig. S1, ESI†), making subsequent analysis of particle morphing impossible. To tackle this problem, we draw multiple lines consecutively with slight overlap. This bundle-up approach appears to strengthen the particles, reduce unwanted distortion, and preserve the designed particle shape. In this work, the studied elastomeric rings are produced by laser-drawing five concentric circles, with each consecutive circle ~2 μm wider in diameter than the previous one. Comparing to those with one-circle-photopolymerization designs, the five-circle-photopolymerized rings maintained their as-defined shape over time, albeit some small change in their aspect ratio still occurred (Fig. S1, ESI†). However, the rings photopolymerized with the five-circle approach are not completely free from shape changing and, in fact, weak anisotropic swelling is still observed (Fig. 2b,c). Immediately after printing, the particles first undergo swelling over a period of about 3,000 s when the size of the particle grows in all directions before saturating. Noticeably, the rate and amount of swelling is greater in the direction perpendicular to the far-field director orientation $\mathbf{n}_0$ (defined as $x$ in the coordinate system) compared to that measured along $\mathbf{n}_0$. Such swelling originates from unpolymerized monomers in the surroundings entering the loosely crosslinked network of the printed elastomers. The overall orientation of the nematic molecular ordering, $\mathbf{n}_0$, breaks the symmetry, with swelling perpendicular to $\mathbf{n}_0$ becoming more pronounced [55]. Therefore, to obtain close-to-round rings, the aspect ratio of the as-printed particles is adjusted to compensate for the anisotropic post-printing swelling. An initial aspect ratio of $d_x:d_y$=1:1.2 ($d_x$, $d_y$ are the length of the outer diameters of the rings measured in *x* and *y* direction) is used in the 3D printing computer programs to generate rings with aspect ratios close to 1 after post-polymerization swelling. While a colloidal ring with initial aspect ratio of 1 would end up being elliptical (Fig. S1, ESI†), this post-printing change of shape can be accounted for in manufacturing both circular and elliptical colloidal particles.

Using the approaches described above, microparticles in the form of rings could be two-photon photopolymerized in the LC bulk of the unpolymerized mesogenic precursors (Fig. 3), as well as when adhered to surfaces (Fig. 4). In the former case, the polymerization directly yields nematic colloids within an aligned LC bulk, whereas the surface-attached microstructures could be used as conventional or nematic colloids with the possibility of additional surface treatment and other modifications. The 3D-photopolymerized colloidal rings exhibit temperature-induced actuation similar to that of large-scale bulk LCE structures (Fig. 3) [64]. When heated above the nematic-isotropic transition temperature, the rings expand in the direction perpendicular to $\mathbf{n}_0$ (defined for the nematic medium before polymerization) and contract along it due to the disappearance of the nematic order. The particles initially have almost a perfect round shape (diameter~30 μm, aspect ratio ~1), but end up becoming elliptical (with axes of the ellipse measured to be 43 μm and 25 μm in *x* and *y* direction, aspect ratio >1.6). The maximum value of $L/L_0$ is 1.43 along $\mathbf{n}_0$, comparable to the typical value 1.5±0.1 found for monodomain bulk LCEs [56]. The fact that the mixture consists of multiple compounds results in a broad biphasic temperature range (50°C ~ 60°C) instead of a single point where the phase transition of the entire system happens. The surrounding LC fluid first turns into the isotropic state at a lower temperature and then the LCE rings follow the order-disorder transition, as evident from the gradual decrease of the birefringence under polarizing microscopy (Fig. 3b). Shape change of the rings in this process may involve contribution from monomers penetrating and staying in the polymer network. To separate this contribution from and evaluate the effect of only temperature, we let the particles swell first until they reach the maximum size without external heat. In this case, when the temperature is increased the rate of shape change is smaller before the phase transition, but in the end the rings still approach similar size and aspect ratio (compare Fig. 3c and 3a). This thermal actuation is reversible as particles return to their previous shape and size when the system is cooled down (Fig. 3d).

### 3.2 Change of surface topology

The LCE colloidal rings can be isolated by printing on the glass slide and washing away the unpolymerized monomer mixture. The particles come out with the designed shape and size and without the swelling observed in the unpolymerized mixture. Although one side of a particle resides on the glass surface, the majority of the particle's volume above the surface can morph freely. Therefore, when heated up above the nematic-isotropic transition temperature, anisotropic actuation is observed: circular rings turn into elliptical ones and vice versa given proper initial aspect ratio (Fig. 4). In a typical case, the aspect ratio is doubled from 1.2 initially to 2.5 in the end (Fig. 4a, b). In contrast, rings printed in a homeotropic cell where the mesogen's director alignment is perpendicular to the glass substrates simply expand radially (Fig. 5). The rings become larger, but their circular cross-section and aspect ratio defined in this geometry remained the same after heating. Although the thermal response of the two kinds of rings appear differently, the physical origins are the same: LCE structures contract along and expand perpendicular to the local director (everywhere roughly along $\mathbf{n}_0$) when transitioning from nematic to isotropic phase.

The drastic change of shape can be engineered to change even the surface topology. To minimize the particles' contact with the surface of the glass slide, we first fabricate scaffolds consisting of parallel lines tilting away from the glass surface using isotropic polymers. Rings with small central holes are then printed by connecting the end tips of the lines, thus supported over the substrate. When heated above the phase transition temperature, the initially closed central holes of the rings open up, effectively changing the topology of the particles from genus=0 to genus=1 (Fig. 4d, e). Similarly-designed homeotropic rings can also change their surface topology in the same way (Fig. 5, bottom right corner of each panel). These processes can be reversed and repeated by thermal cycling, and are a manifestation of the anisotropic actuation capability of LCEs.

The significant actuation and birefringence of the LCE rings indicates that the nematic order is well preserved on the colloidal particle scale (Fig. 3, 4). Previous reports on fabricating monodomain LCE microstructures by soft lithography or microfluidics [56,60] did not pursue the control of particle shapes, while 3D printing has promised higher resolution of defining shape and a larger variety of particle shapes [53,55]. Our work here demonstrates colloidal LCE particles fabricated by 3D printing with the shape-changing capability and facile responses to thermal stimuli comparable to that of bulk LCE materials. When introduced in an ordered fluid such as a nematic LC, colloidal particles may disturb the director field of the host medium and induce topological defects within it, which in turn results in the elasticity-mediated colloidal interactions and colloidal self-assembly [6, 15, 16, 21-28]. For strong and homogeneous surface boundary conditions, the topological characteristics of these defects are in fact dictated by the topology of the particles and, therefore, changing the particle shape and topology, hypothetically, can in turn control the generation and distribution of topological defects in a LC director field [50, 61]. Our shape-morphing and topology-changing particles may provide a way to experimentally observe the dynamics of defect generation and annihilation, such as from loops of line defects to point defects and vice versa. Importantly, since the Euler characteristics and genus of nematic colloidal surfaces are directly related to the topological charges of defects they induce [23, 24], one can expect generation of defects and change of their topological invariants when the topology of the colloidal particles is varied. Such topological transformations would require uniform tangential, perpendicular or conical boundary conditions for the director of the surrounding LC host medium on the particles' surfaces, the pursuing of which is outside the scope of this present study. However, yet another interesting possibility is revealed by our experimental observations in Fig. 3. When colloidal particles are fabricated directly within the nematic host medium of mesogenic precursors (Fig. 3), the situation is different from all the known cases for nematic colloidal objects with tangential, conical and homeotropic boundary conditions [6,23,24,65]. The director field within the photopolymerized part of the system, the LC elastomeric colloidal ring, smoothly transforms into that of the surrounding LC host medium of the unpolymerized monomers without the singular defect formation. This is different from colloidal rings with tangential and homeotropic surface anchoring that both have energy-minimizing structures involving self-compensating surface or bulk defects [23,24] (one should note, however, that these defects around genus-one particles in such conventional nematic colloids appear due to energetic reasons rather than being required by topology because Euler characteristic of such surfaces equals zero).

### 3.3 Reconfiguration of colloidal crystal lattices

To study the collective behavior of the thermal-stimulus-responsive rings, we use the 3D fabrication of particles within the nematic bulk discussed above to produce assemblies of colloidal rings initially photopolymerized in spatial locations corresponding to different 2D lattices (Fig. 6). Constructed away from the confining substrates, the particles are thus free to morph, move, and interact with their neighbors, including when these interactions are mediated by the LC elasticity. Heating these arrays of elastomeric rings above the phase transition temperature change the shape and size of individual particles, i.e. the repeating units, at the lattice points, which, in turn, lead to changes of the overall lattice structure. Among all types of lattices, close-packed ones appear to exhibit the most drastic changes during such processes, which can be understood from the many-body interactions of such shape-morphing micro-objects, given the close proximity of these repeating units. For this reason, a close-packed hexagonal 2D lattice is photopolymerized by placing rings close to each other while leaving just enough room in-between to allow for the anisotropic swelling. In the end, we obtain a multi-particle structure suspended in the nematic monomer mixture where the rings can interact with each other freely and are not connected into one piece (Fig. 6a). During heating, the rings still tend to change their shapes anisotropically relative to $\mathbf{n}_0$, but they are also constrained by their lattice surroundings. The collective effects become apparent. The colloidal rings in the center of the lattice end up stretched almost diagonally (neither along nor normal to $\mathbf{n}_0$), which is partially to fill in the little space between particles while maintaining their relative position. In contrast, colloidal rings at the edges, especially in the vertices of the lattice, have not so many near-neighbor constrains and, therefore, change shape similar to that of individual rings under otherwise similar conditions; some of the particles even get pushed off from the periodic grid, creating defects in the lattice. When cooled down, the deformed and dislocated rings often do not return to their original states upon the thermal cycle end and sometimes exhibit even more emergent changes. This is consistent with the notion that such a thermally reconfigurable colloidal system is intrinsically out of equilibrium. While two-photon-absorption based photopolymerization within the bulk of the LC precursor host medium allows for smoothly matching director structures within interior and exterior of the colloidal rings (Fig. 6), with only minimal distortions of the uniform far-field background defined before polymerization, thermal actuation of particles within the arrays does produce elastic deformations because only the soft solid particles are morphing while the far-field fluid LC medium remains largely unchanged, albeit director of the LC in-between the particles has to adaptively respond to director changes within the particles. These elastic deformations likely also play a role in the ensuing reconfiguration of the elastomeric colloidal ring lattices, but the effects are not necessarily reversible because the scale of elastic energies involved can be on the order of 10,000 $k_BT$ [66].

While conventional nematic colloids tend to form linear chains as well as 2D and 3D lattices that depend on the types of elastic multipoles (e.g. dipoles versus quadrupoles versus hexadecapoles, and so on) that they feature [5-10,16], our elastomeric nematic colloids do not have to exhibit well defined elastic multipoles because of being imprinted within a uniformly aligned host medium. However, our analysis above did demonstrate post-fabrication morphing of these colloidal particles, which certainly can cause additional elastic and order parameter perturbations around them. Such perturbations, and free energy minimization associated with them, can serve as a glue that keeps our particles within the lattice during various types of thermal reconfigurations (Fig. 6). Beyond the mechanism of sharing particle-induced deformations to minimize the overall free energy of the host medium, the effective inter-particle adhesion may also arise from different physical mechanisms because the "blurred" interfaces between the elastomeric and fluid LC components of this system may give origins to a complex interplay of colloidal interactions that still remains to be understood.

As a comparison, slight overlapping of the scanned femtosecond laser trajectories when particles are being produced makes a one-piece microstructure of inseparable microring particles (Fig. S2, ESI†). Upon heating, the particles within such a system stay inter-linked in their relative positions, albeit still exhibiting the expected shape change. The difference is that here the constitute particles themselves are varied, which may result in different lattice constants or entirely different lattice types. In lattices where the repeating units are more separated, shape change of individual particles, although limited to be only on the lattice points, progress as defined by the orientation of $\mathbf{n}_0$ (Fig. 6b, c). Additionally, the lattices can also be engineered to change topology by opening and closing the central holes of the rings through temperature induced phase transition (Fig. 7). The assembly of many rings with changeable topology effectively increases the genus number of the structure to a maximum of $N$, where $N$ is the number of the inter-linked rings. Unlike the previous case, this assembly is fabricated atop of the surface of the substrate. These surface topological patterns may further be used to induce and reconfigure defect patterns in LCs and guide the interaction and self-assembly of colloidal inclusions [10-16]. Such thermal reconfigurations (Figs. 6 and 7), along with the femtosecond-laser-enabled photopolymerization-based patterning, open myriads of possibilities of controlling LCs with which they interface. While obtaining well defined lattices of colloidal microparticles in nematic LCs is often a challenge, our approach allows for imprinting, assembling and reconfiguring lattices with hexagonal, square-periodic, and a variety of other symmetries (Fig. 6), which also is of great fundamental and applied interest.

## 4. Discussion

While all our colloidal and other microstructures presented here were obtained by two-photon photopolymerizing them within a uniformly aligned LC medium, this approach can be extended to the cases where surface anchoring photopatterning first pre-defines the ordering and even defects within the LC, as it was done for elastomeric thin films [51,57]. Combining patterning of director and shapes of photopolymerization-enabled 3D microstructures could allow for "imprinting" well-defined mechanical responses of colloids, say, by placing defects within the bulk of particles being fabricated. Morphing particle shapes or inducing director field undulations through pre-defined boundary conditions would also allow one to further explore various forms of elasticity mediated colloidal self-assembly, such as the lock-and-key mechanism [67]. In addition to photopatterning of surface alignment, one could also exploit twist and other deformations that could be easily endowed into such polymerizable LC systems by chiral dopants and other additives. One can envisage also the multi-component self-assembled colloidal systems where elastomeric particles are just one part of the complex colloidal "soup". Reconfiguring them within these more complex multi-component colloidal assemblies could, hypothetically, impart reconfigurations of both structure and function of the ensuing self-assembled LC colloidal materials.

A significant additional material development is needed to fully take advantage of the elastomeric LC colloidal platform that we have introduced. Although here we were able to perform two-photon polymerization with the elastomeric material at room temperature, the samples in some of the LCE compositions nonetheless suffer from instabilities such as phase separation and crystallization due to the nature of the multi-monomer mixture. Future development of the elastomeric LC colloidal materials would benefit from a simpler and more robust LCE composition, more rigid polymer networks still capable of significant actuation, along with the room-temperature stability for the ease of photopolymerization. Introducing other colloidal particles, such as gold nanorods, quantum dots, and magnetic nanoplatelets into elastomeric LC colloids may further enrich the mechanical actuation with optical, electronic and magnetic responses [61, 62]. Besides, two-photon based polymerization could be combined with other means of control defining both composition and particle shapes, such as microfluidics, holographic laser illumination for cross-linking and optical trapping. Besides colloidal rings, our study could be potentially extended to various colloidal knots and links [25,26] and a variety of other geometries and topologies, some of which have been already considered theoretically specifically for LCE colloidal particles [50], potentially defining a new experimental arena for probing the interplay of topologies of surfaces and the nematic director field.

## 5. Conclusions

In summary, we have demonstrated the fabrication of LCE colloidal particles and microstructures with non-trivial topology and reported their anisotropic thermal actuation of shape and change of topology. Based on this fabrication approach, 2D colloidal crystal lattices have been constructed and examined to explore the out-of-equilibrium emergent dynamics and reconfiguration. The combination of the 3D two-photon photopolymerization technique and responsive LCE material open the possibilities for achieving a wide range of reconfigurable particle shapes and lattice types. These rich combinations may also find uses in building tunable photonic devices and in fundamental research on transformations of director fields in nematic colloids with various nontrivial topological characteristics.

**Acknowledgements:** This work was funded by the U.S. Department of Energy, Office of Basic Energy Sciences, Division of Materials Sciences and Engineering, under contract DE-SC0019293 with the University of Colorado at Boulder (Y.Y. and I.I.S.). I.I.S. also acknowledges the hospitality of the Kavli Institute for Theoretical Physics at the University of California, Santa Barbara, where he was working on this article during his extended stay and where his research was supported in part by the U.S. National Science Foundation under Grant No. NSF PHY-1748958.

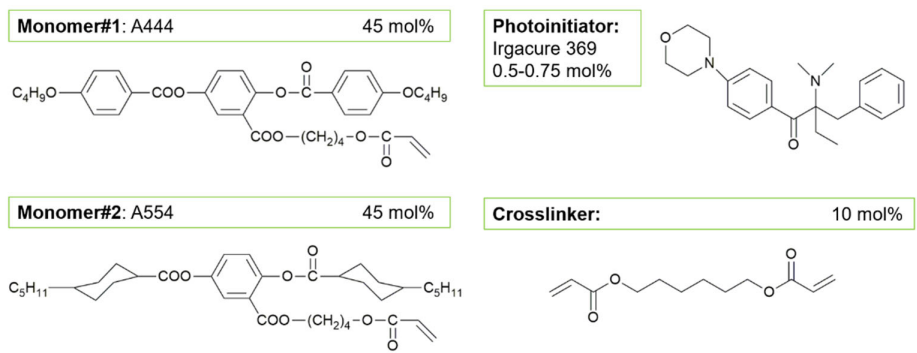

**Fig. 1** Chemical structures of the compounds and their percentages in the mixture used in the experiment. Two monomers, 4"-acryloyloxybutyl 2,5-(4'-butyloxybenzoyloxy)benzoate (A444) and 4"-acryloyloxybutyl 2,5-di(4'-pentylcyclohexyloyloxy)benzoate (A554), of equal molar percentage are used, which allows for the formation of a mixture with a nematic phase at room temperature (25 °C). The rest of the mixture consists of the crosslinker 1,6-hexanediol diacrylate at 10 mol% and the trace amount of photoinitiator, 2-benzyl-2-(dimethylamino)-4'-morpholinobutyrophenone (Irgacure 369).

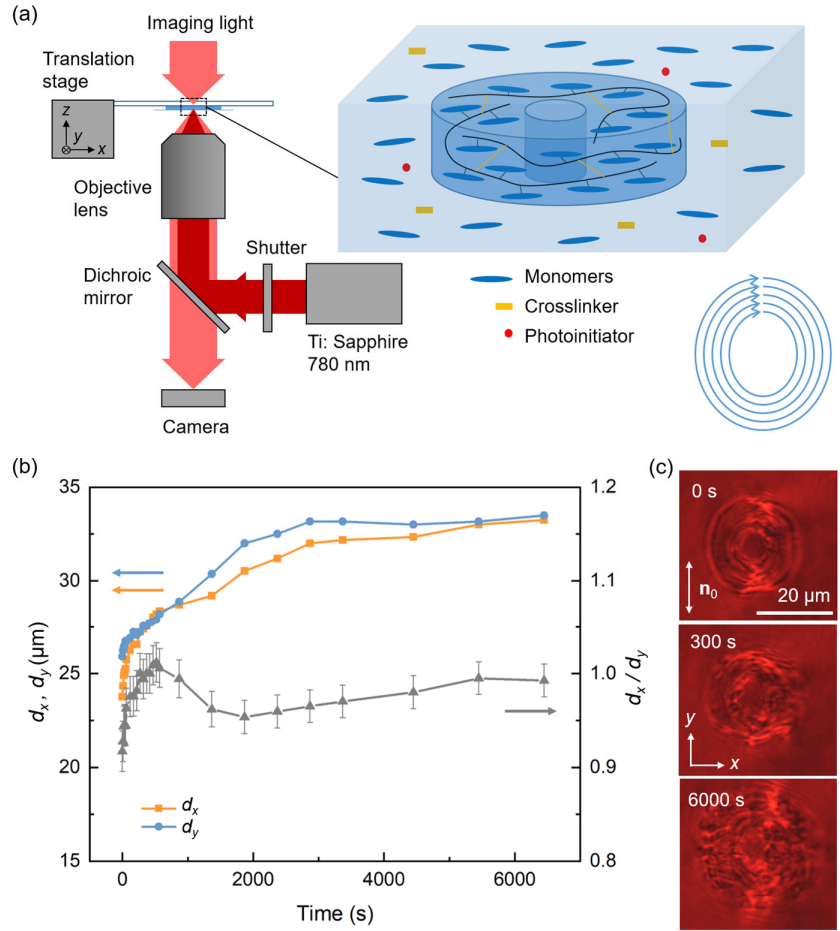

**Fig. 2** Experimental setup used to two-photon photopolymerize elastomeric microstructures. (a) Schematic for the 3D printing setup based on two-photon photopolymerization using a tunable femtosecond laser source. The shutter and the nanopositioning translation stage are controlled by computer programs to cooperatively define the shape of the polymerization-defined colloidal structures. An example of the laser writing trajectories is shown as an inset on the bottom-right corner. (b) Anisotropic swelling of microrings over time. The swelling happens immediately after printing (elapsed time of 0 s) and gradually slows down, which stabilizes after elapsed time of about 5000 s. The coordinate system is constructed such that the far-field director $\mathbf{n}_0$ is parallel to the unit vector $y$, as shown in panel (c). The yellow square symbols and blue dots show the length of the outer diameters of the rings measured in $x$ and $y$ direction, $d_x$ and $d_y$; the grey triangular symbols with error bars show data for the aspect ratio $d_x/d_y$. The measurement errors of $d_x$ and $d_y$ are estimated to be ±0.4 μm to account for the uncertainty from the optical resolution limit and counting number of pixels on the micrograph during the image analysis; these errors are then propagated to obtain the error of the ratio $d_x/d_y$. (c) Snapshots of the swelling process at elapsed times of 0, 300, and 6000 s. White double arrow indicates the far-field director $\mathbf{n}_0$ of the LC host.

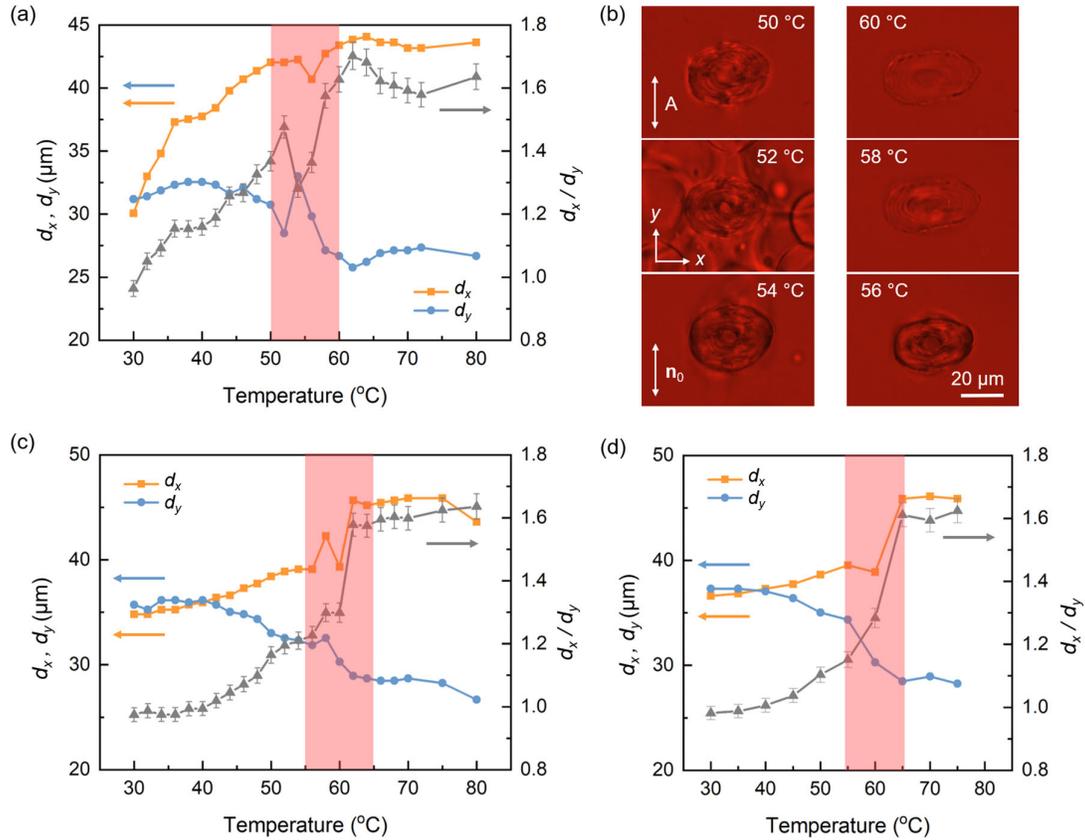

**Fig. 3** Thermal actuation of elastomeric colloidal microstructures within the monomeric LC host. (a) Change of diameters $d_x$, $d_y$, and the ratio $d_x/d_y$ of a microring as temperature increases. Shaded area is where the LC host and the LCE microsctructures subsequently go through the nematic-isotropic phase transition. (b) Snapshots of the system during the phase transition. The LC host first undergoes the transition at ~52 °C and then the printed particles at ~58 °C. White double arrows indicate the orientation of the analyzer (marked "A") and the far-field director $\mathbf{n}_0$ of the LC host. (c) Change of diameters $d_x$, $d_y$, and the ratio $d_x/d_y$ of a microring as temperature increases. The measured particle is first allowed to swell until reaching the equilibrium state. We note that the different kinetic trajectories (with and without pre-swelling before heating up) of the particles shown in (a) and (c) have led to slightly different biphasic regions and shape morphing behaviors. (d) Change of diameters $d_x$, $d_y$, and $d_x/d_y$ as temperature decreases, using the same particle as in (c).

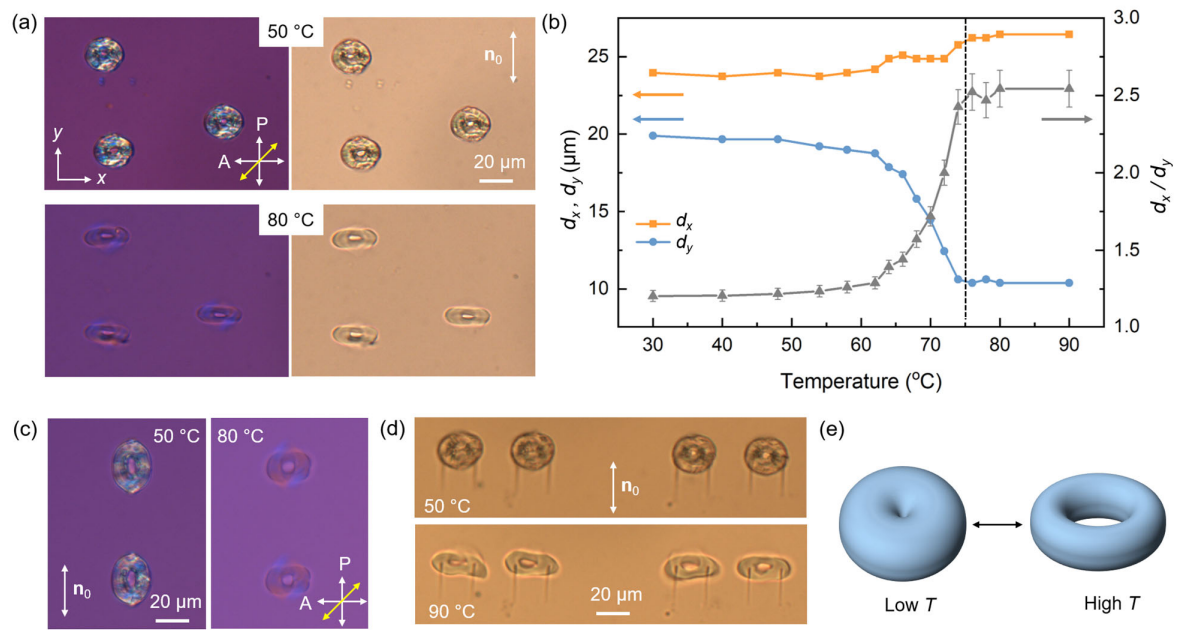

**Fig. 4** Surface-attached colloidal LCE structures. (a) Micrographs of microrings on the surface of a glass substrate before and after phase transition. Pictures on the left are taken under crossed polarizers (white double arrows marked with "A", "P") with a 530 nm retardation plate (yellow double arrow indicating the slow axis). White double arrow indicates the far-field director $\mathbf{n}_0$ during laser writing. (b) Change of diameters $d_x$, $d_y$, and $d_x/d_y$ of a surface-attached microring as temperature increases. The phase transition finishes at ~75 °C, after which actuation of the microrings stops. (c) Micrographs of surface ellipses before and after phase transition. (d) Change of topology of microrings elevated from the glass substrate. The two supporting sticks for each particle are 3D-printed lines that tilt up from the substrates. After phase transition, holes appear at the center of the two microrings on the left of the images, making them genus=1. (e) 3D schematic showing how the change of topology happens in (d), with the hole of the 2D toroidal surface of the colloidal particle opening/closing upon thermal actuation.

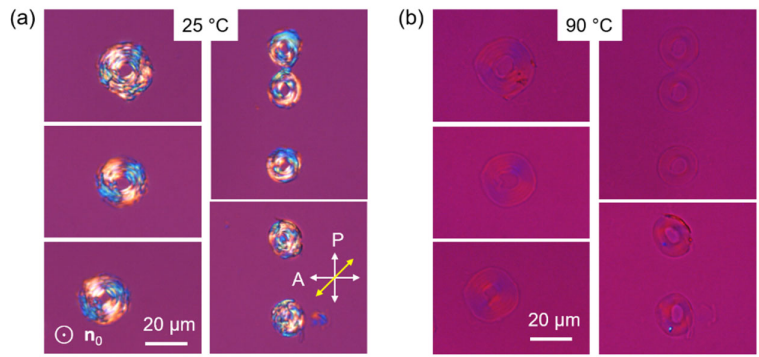

**Fig. 5** Surface-adhered LCE microrings made in homeotropic LC cells. In contrast to the ones in Fig. 4, shape change of these microrings is axially symmetric and isotropic in the viewing plane, though change of topology from genus-zero to genus-one is also observed, as shown in the bottom left corner. Far-field director $\mathbf{n}_0$ during printing is perpendicular to the viewing plane as indicated by the circled dot symbol.

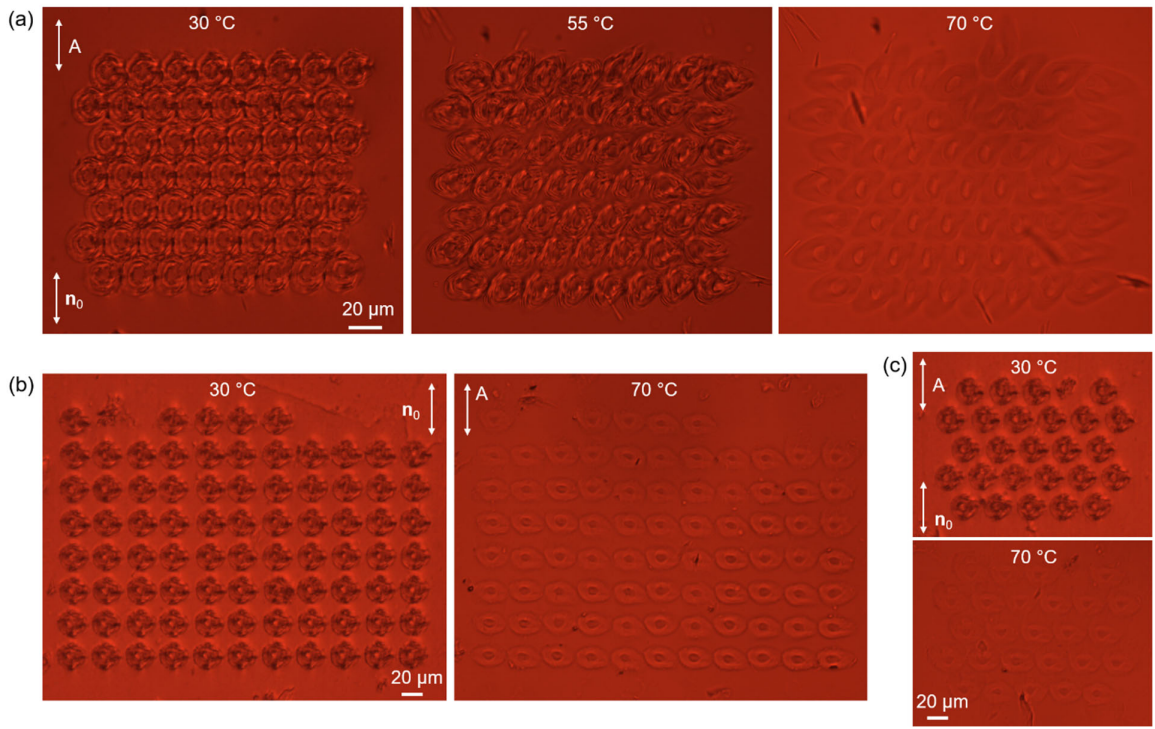

**Fig. 6** Elastomeric colloidal crystal lattices in a LC host. Micrographs are taken under different temperatures for originally close-packed hexagonal (a), sparse-packed square (b), and sparse-packed hexagonal (c) lattices. The repeating units are the microrings suspended in the LC host and not connected to each other, so that the particles are free to move around.

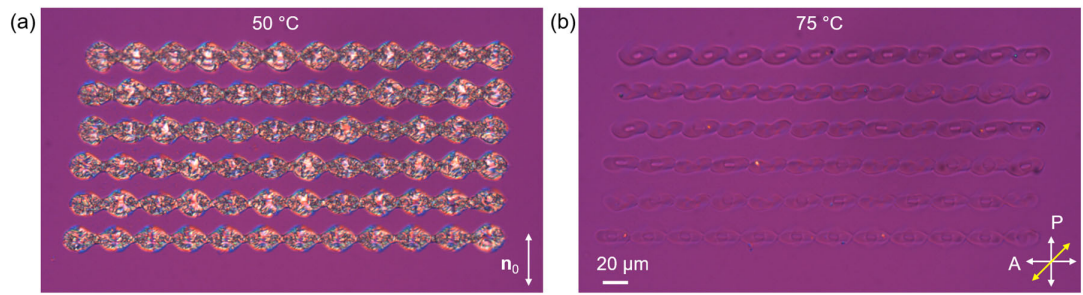

**Fig. 7** Surface-attached LCE microstructures. A lattice made of elastomeric microrings is constructed on the surface of a glass substrate. The sample (a) is heated above the phase transition temperature (b) of the polymerized LCE such that actuation of individual particles and changes of topology are observed.

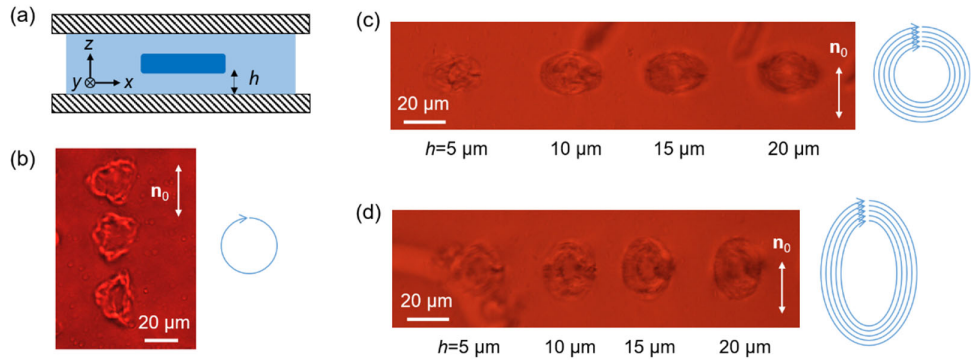

**Fig. S1** Fabrication of free-standing colloidal elastomeric microrings within a LC cell. (a) Schematic showing how the printed particles are placed in the middle of the cell, with a distance $h$ from the bottom substrate. The particle is represented by the blue rectangle, the polymerizable LC host by the light blue background, and the glass substrates by the shaded rectangles. (b) Micrographs of particles consisting of only one ring. The shape cannot be maintained because of distorted polymerized structure from swelling. (c, d) To strengthen the particles and thus maintain their shape, particles fabricated by guiding the focused laser beam along five concentric circles (c) or ellipses (d) are printed. Varying $h$ up to 50 micrometers does not result in observable difference in particle shape.

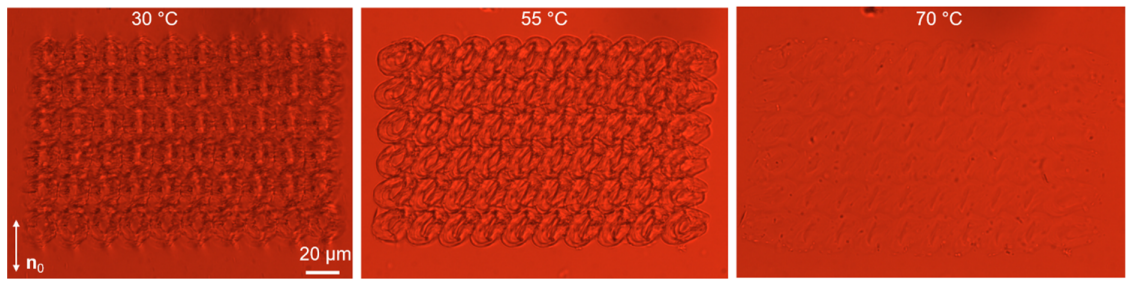

**Fig. S2** Thermal actuation of LCE microstructures in the LC host. Microrings are placed following the pattern of square lattices but with slight overlap. This resulted in a structure where the repeating units are not freely moving but instead, bound together into a one-piece microstructure.